\documentclass[12pt]{article}
\usepackage{verbatim}
\usepackage{amsfonts}
\usepackage{graphics}
\usepackage{amsmath}
\usepackage{times}
\usepackage{appendix}
\usepackage{color}
\usepackage{soul}
\usepackage{mathtools}
\usepackage{enumerate}
\usepackage{fancyhdr,latexsym,amsmath,amsfonts,amssymb,amsbsy,amsthm,url}
\usepackage{amsthm,amssymb,mathrsfs,setspace,pstcol}
\usepackage[margin=0.5in,footskip=0.25in]{geometry}
\usepackage{graphics,graphicx,epsfig}
\usepackage{breqn}
\usepackage{caption,subcaption,float,subfloat}
\usepackage{pdflscape}
\usepackage{hyperref}
\usepackage[ruled,vlined]{algorithm2e}

\usepackage[english]{babel}
\usepackage[dvipsnames]{xcolor}
\makeatletter

\renewcommand{\section}{
	\@startsection
	{section}
	{1}
	{0pt}
	{1.1\baselineskip}
	{0.2\baselineskip}
	{\sc \centering}
}

\renewcommand{\subsection}{
	\@startsection
	{subsection}
	{1}
	{0pt}
	{1.1\baselineskip}
	{0.2\baselineskip}
	{\sc \centering}
}

\renewcommand{\subsubsection}{
	\@startsection
	{subsubsection}
	{1}
	{0pt}
	{1.1\baselineskip}
	{0.2\baselineskip}
	{\sc \centering}
}

\makeatother

\usepackage[flushleft]{threeparttable}
\usepackage{rotating,booktabs,multirow}
\usepackage{colortbl}
\usepackage{makecell,cellspace,caption}

\begin{document}
	
\title{\large\sc Classification based credit risk analysis: The case of Lending Club}
\normalsize
\author{\sc{Aadi Gupta} \thanks{Department of Mathematics, Indian Institute of Technology Guwahati, Guwahati-781039, India, e-mail: aadi18@alumni.iitg.ac.in}
\and 
\sc{Priya Gulati} \thanks{Department of Mathematics, Indian Institute of Technology Guwahati, Guwahati-781039, India, e-mail: gulati18@alumni.iitg.ac.in}
\and 
\sc{Siddhartha P. Chakrabarty} \thanks{Department of Mathematics, Indian Institute of Technology Guwahati, Guwahati-781039, India, e-mail: pratim@iitg.ac.in}
}

\date{}
\maketitle
\begin{abstract}

In this paper, we performs a credit risk analysis, on  the data of past loan applicants of a company named Lending Club. The calculation required the use of exploratory data analysis and machine learning classification algorithms, namely, Logistic Regression and Random Forest Algorithm. We further used the calculated probability of default to design a credit derivative based on the idea of a Credit Default Swap, to hedge against an event of default. The results on the test set are presented using various performance measures.

{\it Keywords: Credit risk; Classification algorithm; Exploratory data analysis}

\end{abstract}

\section{Introduction}
\label{Sec_Introduction}

Lending Club, headquartered in San Francisco, was the first peer-to-peer lending institution to offer its securities through the Securities and Exchange Commission (SEC) and enter the secondary market \cite{wikilendingclub}. This article is essentially a case study, on how financial engineering problems can be addressed using Machine Learning (ML) and Exploratory Data Analysis (EDA) approaches. Lending Club specializes in extending different types of loans to urban customers, which is decided on the basis of the applicant's profile. Accordingly, the data considered in this work, contains information about past loan applications and whether they were ``defaulted'' or ``not''. One of the main objectives of this work is the calculation of the Probability of Default (PD) and (in case of a default) the determination of  Loss Given Default (LGD), Exposure at Default (EAD), and finally, the Expected Loss (EL), making use of the historical data. Also, using the ``Recovery Rate'' (estimated while calculating ``LGD'' from ``EAD'') and the PD, a simple credit derivative is implemented, based on the concept of Credit Default Swaps (CDS) which can be used to hedge against such defaults. It may be noted that the total value a lender is exposed to when a loan defaults, is the EAD and the consequent unrecoverable amount for the lender is the LGD \cite{bluhm16}. Accordingly, EL is defined as,
\[\text{EL}=\text{EAD}\times\text{LGD}\times\text{PD}.\]
The driver of this work is the notion of ``Classification Algorithm'', which weighs the input data (of the applicant, in this case) to classify the input features into positive and negative classes \cite{kim20} (default on the loan or not, in this case). Accordingly, we consider two ``Classification Algorithms'', namely, the Logistic Regression and the Random Forest. 

The main idea behind Logistic Regression is to determine the probability of a particular data point belonging to the positive class (in the case of binary classification) \cite{peng02}. The model does so by establishing a ``linear relationship'' between the independent and the dependent variables. The weights of the linear relations is determined through the minimization of a Loss Function, which is achieved by using the ``Gradient Descent Optimization Algorithm''.  Since Logistic Regression first predicts the probability of belonging to the positive class, therefore it creates a linear decision boundary (based on a \textit{threshold}, set by the user), separating the two classes from one another. This decision boundary can now be represented as a conditional probability \cite{sperandei14}. Implementation of the Random Forest algorithm involves the training stage construction of several decision trees \cite{ali12}, and predictions emanating from these trees are averaged to arrive at a final prediction. Since the algorithm uses an average of results to make the final prediction, the Random Forest algorithm is referred to as an ensemble technique. Decision Trees are designed to optimally split the considered dataset into smaller and smaller subsets, in order to predict the value being targeted \cite{patel18,sharma16}. Some of the criterion used to calculate the purity or impurity of a node, include Entropy and Gini Impurity. In summary, a decision tree splits the nodes on all the attributes present in the data, and then chooses the split with the most Information Gain, with the Decision Tree model of Classification and Regression Trees (CART), being used in this paper.

The firm based model uses the value of a firm to represent the event of default, with the default event being represented by the boundary conditions of the process and the dynamics of the firm value \cite{duffie99}. In particular, we refer to two well-established models. Firstly, we mention the Merton model based on the the seminal paper of Black and Scholes \cite{schoutens10}, which is used to calculate the default probability of a reference entity. In the context, the joint density function for the first hitting times is determined \cite{shreve04}. Secondly, we have the Black-Cox model, which addresses some of the disadvantages of the Merton model \cite{schoutens10}. In order to hedge against credit risk, the usage of credit derivatives is a customary approach \cite{garcia07,bielecki13,schonbucher03}, with Credit Default Swaps (CDS) being the most common choice of credit derivatives. This type of contracts entail the buyer of the CDS to transfer the credit risk of a reference entity (``Loans'' in this case constitute the reference entity) to the seller of the protection, until the credit has been settled. In return, the protection buyer pays premiums (predetermined payments) to the protection seller, which continues until the maturity of the CDS or a default, whichever is earlier \cite{schoutens10}. The interested reader may refer to \cite{schoutens10} for the formula of CDS spread per annum, to be used later in this paper. Another widely used credit derivative, albeit more sophisticated than the CDS are the Collateralized Debt Obligation (CDO), which is a structured product, based on tranches \cite{schonbucher03}. CDOs can further classified into cash, synthetic and hybrid. The interested reader may refer to \cite{schonbucher03} for a detailed presentation on pricing of synthetic CDOs.

\section{Methodology} 

The goal of this exercise is the approximation of a classification model, on the data considered (from the peer-to-peer lending company Lending Club), in order to predict as to whether an applicant (whose details are contained in the considered database) is likely to default on the loan or not. Accordingly, to this end, it is necessary to identify and understand the essential variables, and take into account the summary statistics, in conjunction with data visualization. The dataset used for the estimation of PD, EAD, LGD and EL was obtained from Kaggle \cite{lendingclub}. The data contains details of all the applicants who had applied for a loan at Lending Club. There were separate files obtained for the accepted and the rejected loans. The file ``accepted loans'' was only used, since the observations were made on the applicants who ultimately paid the loan, and those who defaulted on the loan. The data involved the details of the applicants for the loan, such as FICO score, loan amount, interest rate, purpose of loan etc. Machine Learning (ML) algorithms were applied on this data to predict the PD after data exploration, data pre-processing and feature cleaning.

The feature selection was performed on the considered data, bearing in mind the goals of predictive modelling. Given the large number of existent features, it was desirable to achieve a reduction in their numbers, thereby retaining only the relevant ones. This action abetted the reduction in computational cost and improvement in the performance of the model.

The data pre-processing was carried out before the application of the classification model, in order to obtain the probabilities of default. This exercise involved the removal or filling of missing data, removing features which are repetitive or deemed unnecessary and the conversion of  categorical features to dummy variables. Some of the dropped columns were ``emp\_title'', ``emp\_length'', ``grade'', ``issue\_d'' and ``title''. After pre-processing, the data was divided into the customary training set and test set. The former was then used to train the classification model, with the results being evaluated subsequently, on the latter.

In order to accomplish the goal of comparison of different models, the problem of predicting the PD (denoted by $p$) was treated as a classification problem, where, if $p \geq 0.5$, it is treated as $1$ (\textit{i.e.,} the person will default) and if $p < 0.5$, it is treated as $0$ (\textit{i.e.,} the person will not default). Accordingly, $p=0.5$ was used as the \textit{threshold} for the classifier.

It was observed from the data, that the classes are imbalanced, and therefore, accuracy is not a good metric, when it comes to comparison of different models. Accordingly, we instead used the ``Confusion Matrix'' (to evaluate the performance of our model), which is a table with four different combinations of actual and predicted values \textit{i.e.,} True Positive (TP), True Negative (TN), False Positive (FP) and False Negative (FN). It is beneficial for measuring \textit{Recall} $\displaystyle{\left(\frac{\text{TP}}{\text{TP+FN}}\right)}$, \textit{Precision} $\displaystyle{\left(\frac{\text{TP}}{\text{TP+FP}}\right)}$ and \textit{F1-score} (Harmonic Mean of \text{Recall} and \text{Precision}).

Receiver Operating Characteristic (ROC) curve is used to compare the performance of a classification algorithm by varying the threshold value \cite{sonego08}. ROC is a probability curve and the area under the curve (AUC) represents the likelihood (Random Positive has dominant ranking over Random Negative) \textit{i.e.,} the degree of separability. The AUC represents how successful is the model in differentiating between the classes. The higher the AUC value, the more amenable is the model at classifying $0$ classes as $0$, and $1$ classes as $1$ (\textit{i.e.,} correct prediction). The ROC curve was plotted with TPR (True Positive Rate) against the FPR (False Positive Rate) where FPR is on the $x$-axis and TPR is on the $y$-axis. Note that TPR is the same as \textit{Sensitivity} and $\displaystyle{\text{1-FPR}=\frac{\text{TN}}{\text{FP+TN}}}$ is also called \textit{Specificity}.

\section{Results}

\begin{figure}[h]
\centering
\includegraphics[scale=0.75]{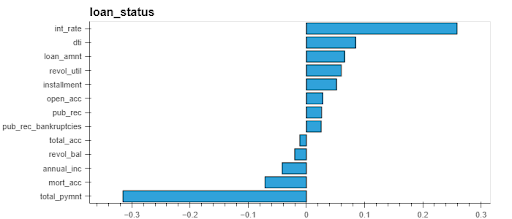}
\caption{Correlation Coefficient between Target Variable and Various Attributes}
\label{Loan_Status}
\end{figure}

The statistics-based feature selection methods were used to assess the correlation between every input variable and the target variable, and those input variables which demonstrated the strongest relationship vis-a-vis the target variable, were selected. Further, for selecting the relevant input variables, a comparative analysis was performed and some of the observations, as a result of this exercise, are noted below:
\begin{enumerate}[(A)]
\item It looks like $F$ and $G$ sub-grades do not get paid back that often. 
\item The loans with the highest interest rates are faced with a greater likelihood of default
\item Only $339$ borrowers have reported an annual income exceeding $1$ million.
\item It seems that the more the ``dti'', the more likely is it that the loan will not be paid.
\item Only $1211$ borrowers have more than $40$ open credit lines.
\item Only $1299$ borrowers have more than $80$ credit lines in the borrower credit file.
\end{enumerate}
The graph in Figure \ref{Loan_Status} shows the correlation between various features, like int\_rate and loan\_status (\textit{i.e.,} whether the debtor will default or not). Positive correlation implies that more the value of  particular feature, more is the chance that the debtor will default on their loan.

\subsection{Logistic Regression}

The classification report displays precision, recall and f1-score for both the classes $0$ and $1$ \textit{i.e.,} applicants who will not default and applicants who will default, respectively. Apart from that, the report also shows Macro Average ($0.5\times \text{score}_{0}+0.5\times \text{score}_{1}$) and the weighted average ($w_{0}\times \text{score}_{0}+w_{1}\times \text{score}_{1}$, where $w_{0}$ and $w_{1}$ are set according to the number of data points in each class). Accordingly, the Accuracy Score obtained is $97.17\%$ and the Classification Report for the ``Train Result'' is given in Table \ref{Tab_03_One}. 
\begin{table}[h]
\centering
\begin{tabular}{c|ccccc}
\hline\hline
& 0.0 &  1.0 & Accuracy & Macro Average & Weighted Average\\ 
\hline
Precision & 0.97 & 0.99 &  0.97 &  0.98 & 0.97 \\
Recall & 1.00 & 0.87 & 0.97 & 0.93 & 0.97  \\
f1-score & 0.98 & 0.92 & 0.97 & 0.95 & 0.97 \\
\hline\hline
\end{tabular}
\caption{Train Evaluation Metrics for Logistic Regression}
\label{Tab_03_One}
\end{table}

Further, the Accuracy Score is $88.81\%$ and the Classification Report for the ``Test Result'' is given in Table \ref{Tab_03_Two}.
\begin{table}[h]
\centering
\begin{tabular}{c|ccccc}
\hline\hline 
& 0.0 & 1.0 & Accuracy & Macro Average & Weighted Average \\ 
\hline
Precision & 0.98 & 0.65 & 0.89 & 0.82 & 0.92 \\
Recall & 0.87 & 0.95 & 0.89 & 0.91 & 0.89  \\
f1-score & 0.93 & 0.77 & 0.89 & 0.85 & 0.90 \\
\hline\hline
\end{tabular}
\caption{Test Evaluation Metrics for Logistic Regression}
\label{Tab_03_Two}
\end{table}

\begin{figure}[H]
\centering
\includegraphics[scale=0.6]{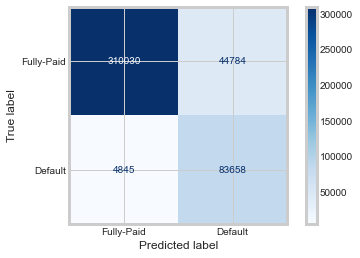}
\caption{Confusion Matrix for Logistic Regression Model}
\label{cmLR}
\end{figure}

\begin{figure}[H]
\centering
\includegraphics[scale=0.6]{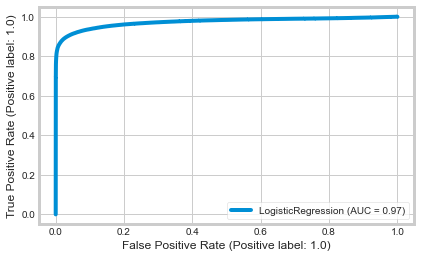}
\caption{ROC Curve for Logistic Regression Model}
\label{aucLR}
\end{figure}

The Confusion Matrix for Logistic Regression Model and the ROC Curve for Logistic Regression Model are given in Figures \ref{cmLR} and \ref{aucLR}, respectively.

\subsection{Random Forest Classifier}

For the Random Forest classifier, the Accuracy Score obtained is $100.00\%$ and the Classification Report for the ``Train Result'' is given in Table \ref{Tab_03_Three}.
\begin{table}[h]
\centering
\begin{tabular}{c|ccccc}
\hline\hline	
& 0.0 & 1.0 & Accuracy & Macro Average & Weighted Average \\ \hline
Precision & 1.00 & 1.00 & 1.00 & 1.00 & 1.00 \\
Recall & 1.00 & 1.00 & 1.00 & 1.00 & 1.00 \\
f1-score & 1.00 & 1.00 & 1.00 & 1.00 & 1.00 \\
\hline\hline
\end{tabular}
\caption{Train Evaluation Metrics for Random Forest}
\label{Tab_03_Three}
\end{table}

Further, the Accuracy Score is $97.11\%$ and the Classification Report for the ``Train Result'' is given in Table \ref{Tab_03_Four}.
\begin{table}[h]
\centering
\begin{tabular}{c|ccccc}
\hline\hline 
& 0.0 & 1.0 & Accuracy & Macro Average & Weighted Average \\ \hline
Precision & 0.97 & 1.00 & 0.97 & 0.98 & 0.97 \\
Recall & 1.00 & 0.86 & 0.97 & 0.93 & 0.97 \\
f1-score & 0.98 & 0.92 & 0.97 & 0.95 & 0.97 \\
\hline\hline
\end{tabular}
\caption{Test Evaluation Metrics for Random Forest}
\label{Tab_03_Four}
\end{table}

\begin{figure}[H]
\centering
\includegraphics[scale=0.6]{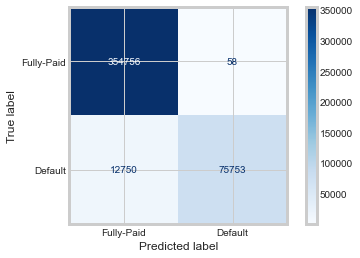}
\caption{Confusion Matrix for Random Forest Model}
\label{cmRF}
\end{figure}

\begin{figure}[H]
\centering
\includegraphics[scale=0.6]{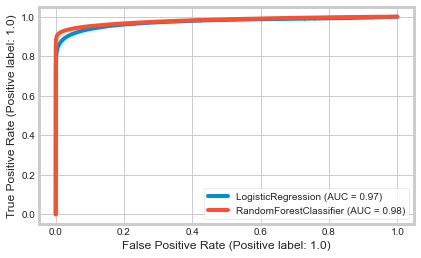}
\caption{ROC Curve for Random Forest Model}
\label{aucRF}
\end{figure}

The Confusion Matrix for Logistic Regression Model and the ROC Curve for Logistic Regression Model are given in Figures \ref{cmRF} and \ref{aucRF}, respectively

\subsection{Comparison}

\begin{figure}[H]
\centering
\includegraphics[scale=0.8]{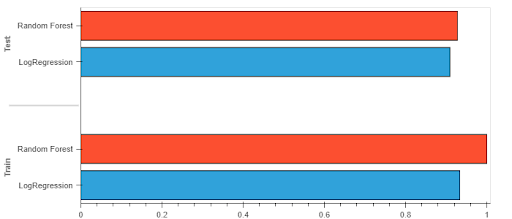}
\caption{Graph showing Train and Test Accuracy for both the models}
\label{Fig_03_Comparison}
\end{figure}

Finally, the comparative analysis, based on the AUC are presented in Table \ref{Tab_03_Comparison} and Figure \ref{Fig_03_Comparison}. It may be noted from  Table \ref{Tab_03_Comparison}, that in terms of the AUC score, Logistic Regression performs better than the Random Forest.
\begin{table}[H]
\centering
\begin{tabular}{c|c}
\hline\hline
& roc\_auc\_score \\ 
\hline
Random Forest & 0.928\\
Logistic Regression & 0.910 \\
\hline\hline
\end{tabular}
\caption{AUC Score for Logistic Regression and Random Forest}
\label{Tab_03_Comparison}
\end{table}

\subsection{Calculating EAD, LGD, EL and price of the Credit Derivative}

We begin with the following assumptions in order to calculate EAD and LGD. \href{https://www.kansascityfed.org/documents/1381/2012-What\%20Determines\%20Creditor\%20Recovery\%20Rate\%3F.pdf}{Recovery rate} (which is used to calculate the LGD and the price of the credit derivative) is calculated using the recoveries mentioned for the  ``Charged Off'' clients present in the data. Also, the recovery rate is calculated for each department separately. Pricing of the credit derivative is based upon the theory highlighted in Section \ref{Sec_Introduction}. Further, the following assumptions are made while pricing the derivative:
\begin{enumerate}[(A)]
\item Only $1$ payment occurs from protection buyer to protection seller (either at maturity or at the time of default).
\item The prevalent (risk-free) interest rates are taken into account while calculating the discounting factor.
\item For the calculation of accural amount,we assume that on an average defaults occur right in between of a payment period.
\item At the time of the contract writing, the remaining loan amount (including interest rate) is taken as the notional amount and the loan is treated as a bond.
\end{enumerate}

Table \ref{Tab_03_EAD_LGD_EL_CD_One} shows some (illustrative) debtors with their probability of default, along with EAD, LGD, EL and the price of the credit derivative.
\begin{table}[h]
\centering
\begin{tabular}{|c|c|c|c|c|}
\hline\hline
Probability of Default & EAD & LGD & EL & Credit Derivative Price (in bps)\\ 
\hline
0.09 & 18330 & 16727.9 & 1505.52 &  0.0469114 \\
0.41 & 11222.7 & 10241.8 & 4199.14 & 0.156411 \\
0.82 & 20403.9 & 18620.6 & 15268.9 & 0.447656 \\ 
0.97 & 37936.2 & 34620.6 & 33581.9 & 0.38918\\
\hline\hline
\end{tabular}
\caption{Table showing EAD, LGD, EL and derivative price for some (illustrative) customers}
\label{Tab_03_EAD_LGD_EL_CD_One}
\end{table}

\section{Conclusion}

The main objective of the paper was the calculation of probability of default for the debtors of a loan lending organization and the consequent design of a credit derivative (\textit{i.e.,} to calculate the premium) thereby hedging against such defaults. Some conclusions based on the results obtained are as follows:
\begin{enumerate}[(A)]
\item Based on the performance metrics, the Decision Tree model performs better than the Logistic Regression model. \item Further, when the dependent and independent variables have a high non-linearity and complex relationship, linear models like Logistic Regression will not perform well, since the entire algorithm is based on finding a linear relationship (line of best fit in Logistic Regression) between the variables. In such cases, a tree model (non-linear models) will outperform a classical regression method. Hence, there exists a non-linear relationship between the various attributes and the calculated probability.
\item As expected, the calculated derivative premium is higher for loans linked with obligors with high probability of default and short term loans.
\end{enumerate}

\bibliographystyle{elsarticle-num}

\bibliography{BIBLIO}
	
\end{document}